\documentclass[]{spie}  %>>> use for US letter paper
\usepackage[]{graphicx}
\usepackage{subfigure}
\usepackage{epsfig}

\title{EBEX: A balloon-borne CMB polarization experiment} 

\author{Britt Reichborn-Kjennerud\supit{a},
Asad M. Aboobaker\supit{b},
Peter Ade\supit{c},
Fran\c cois Aubin\supit{d},
Carlo Baccigalupi\supit{e},
Chaoyun Bao\supit{b},
Julian Borrill\supit{f},
Christopher Cantalupo\supit{f},
Daniel Chapman\supit{a},
Joy Didier\supit{a},
Matt Dobbs\supit{d},
Julien Grain\supit{g},
William Grainger\supit{c},
Shaul Hanany\supit{b},
Seth Hillbrand\supit{a},
Johannes Hubmayr\supit{h},
Andrew Jaffe\supit{i},
Bradley Johnson\supit{j},
Terry Jones\supit{b},
Theodore Kisner\supit{f},
Jeff Klein\supit{b},
Andrei Korotkov\supit{k},
Sam Leach\supit{e},
Adrian Lee\supit{j},
Lorne Levinson\supit{l},
Michele Limon\supit{a},
Kevin MacDermid\supit{d},
Tomotake Matsumura\supit{m},
Xiaofan Meng\supit{j},
Amber Miller\supit{a},
Michael Milligan\supit{b},
Enzo Pascale\supit{c},
Daniel Polsgrove\supit{b},
Nicolas Ponthieu\supit{g},
Kate Raach\supit{b},
Ilan Sagiv\supit{b},
Graeme Smecher\supit{d},
Federico Stivoli\supit{n},
Radek Stompor\supit{o},
Huan Tran\supit{j},
Matthieu Tristram\supit{p},
Gregory S. Tucker\supit{k},
Yury Vinokurov\supit{k},
Amit Yadav\supit{q},
Matias Zaldarriaga\supit{q},
Kyle Zilic\supit{b},
\skiplinehalf
\supit{a}Columbia University, New York, NY 10027;
\supit{b}University of Minnesota School of Physics and Astronomy, Minneapolis, MN 55455;
\supit{c}Cardiff University, Cardiff, CF24 3AA, United Kingdom;
\supit{d}McGill University, Montr\'eal, Quebec, H3A 2T8, Canada;
\supit{e}Scuola Internazionale Superiore di Studi Avanzati, Trieste 34151, Italy;
\supit{f}Lawrence Berkeley National Laboratory, Berkeley, CA 94720;
\supit{g}Institut d'Astrophysique Spatiale, Universite Paris-Sud, Orsay, 91405, France;
\supit{h}National Institute of Standards and Technology, Boulder, CO, 80303;
\supit{i}Imperial College, London, SW7 2AZ, England, United Kingdom;
\supit{j}University of California, Berkeley, Berkeley, CA 94720;
\supit{k}Brown University, Providence, RI 02912;
\supit{l}Weizmann Institute of Science,  Rehovot 76100, Israel;
\supit{m}California Institute of Technology, Pasadena, CA 91125;
\supit{n}Institut National de Recherche en Informatique et Automatique, Universite Paris-Sud, Orsay, 91405, France;
\supit{o}CNRS, Laboratoire Astroparticule et Cosmologie (APC), Universit\'e Paris Diderot, Paris Cedex 13, 75205, France
\supit{p}Laboratoire de l'Acc\'el\'erateur Lin\'eaire, Universit\'e Paris Sud, CNRS, Orsay, France;
\supit{q}Institute for Advanced Study, Princeton, NJ 08540\
}

\authorinfo{Further author information: (Send correspondence to B. Reichborn-Kjennerud)\\ 
B. R.-K.: E-mail: britt@phys.columbia.edu, Telephone: 1 (212) 854-4211}
\begin{document} 
  \maketitle 

%%%%%%%%%%%%%%%%%%%%%%%%%%%%%%%%%%%%%%%%%%%%%%%%%%%%%%%%%%%%% 
\begin{abstract}
EBEX is a NASA-funded balloon-borne experiment designed to measure the polarization of the cosmic microwave background (CMB).  Observations will be made using 1432 transition edge sensor (TES) bolometric detectors read out with frequency multiplexed SQuIDs.  EBEX will observe in three frequency bands centered at 150, 250, and 410 GHz, with 768, 384, and 280 detectors in each band, respectively.  This broad frequency coverage is designed to provide valuable information about polarized foreground signals from dust.  The polarized sky signals will be modulated with an achromatic half wave plate (AHWP) rotating on a superconducting magnetic bearing (SMB) and analyzed with a fixed wire grid polarizer.  EBEX will observe a patch covering $\sim$1\% of the sky with 8' resolution, allowing for observation of the angular power spectrum from $\ell$ = 20 to 1000.  This will allow EBEX to search for both the primordial $B$-mode signal predicted by inflation and the anticipated lensing $B$-mode signal.  Calculations to predict EBEX constraints on r using expected noise levels show that, for a likelihood centered around zero and with negligible foregrounds, 99\% of the area falls below $r = 0.035$.  This value increases by a factor of 1.6 after a process of foreground subtraction.  This estimate does not include systematic uncertainties.  An engineering flight was launched in June, 2009, from Ft. Sumner, NM, and the long duration science flight in Antarctica  is planned for 2011.  These proceedings describe the EBEX instrument and the North American engineering flight.  
\end{abstract}

%>>>> Include a list of keywords after the abstract 


%%%%%%%%%%%%%%%%%%%%%%%%%%%%%%%%%%%%%%%%%%%%%%%%%%%%%%%%%%%%%
\section{INTRODUCTION}
\label{sec:intro}  
%However, the field of cosmology lacks a specific, well-supported description of the dynamics in the early Universe, and a mechanism to generate the seeds of the density perturbations that grew into the Universe we observe today, both of which are provided by various inflationary models.  
The data from a host of experiments support the current standard cosmological model, including an inflationary epoch shortly after the big bang.  Cosmic microwave background (CMB) polarization data is the most promising probe of the inflationary epoch since, in the inflationary paradigm, quantum fluctuations stretched by inflation to astronomical scales imprinted a unique primordial $B$-mode polarization signal in the CMB.  Additionally, measurement of the CMB $B$-mode polarization probes the integrated structure along the line of sight due to lensing of the CMB by matter between the surface of last scattering and the present epoch (Zaldarriaga and Seljak 1998).  Although E-mode polarization has been detected by a number of experiments, neither the primordial inflationary nor the lensing $B$-mode signals have been detected.  
%~\cite{Kovac:2002, Chiang:2010, Brown:2009}

The EBEX optical design, large number of detectors, and scan strategy provide high sensitivity to intermediate and small angular scales to allow for measurement of both the primordial and lensing $B$-mode signals.  These proceedings provide an overview of the EBEX science goals, the instrument, and the North American engineering flight, updating the instrument description presented previously~\cite{Oxley:2004, Grainger:2008}.  Other recent papers provide more details on specific subsystems~\cite{Hubmayr:2008, Aubin:2010, Milligan:2010}.

%%%%%%%%%%%%%%%%%%%%%%%%%%%%%%%%%%%%%%%%%%%%%%%%%%%%%%%%%%%%%
\section{SCIENCE GOALS}
%While the lensing B-mode signal is weak on large scales, it is comparable in amplitude to the gravitational wave B-mode on intermediate scales and it dominates the B-mode signal at small scales.  The shape of the lensing signal mirrors that of the small scale E-mode signal, where the acoustic peak structure is smeared out.

\subsection{Probe the Inflationary Epoch}
\label{sec:probeinfl}
\begin{figure}
\centering
\subfigure[] % caption for subfigure
{
    \label{sci:sub:a}
    \includegraphics[height=2.6in]{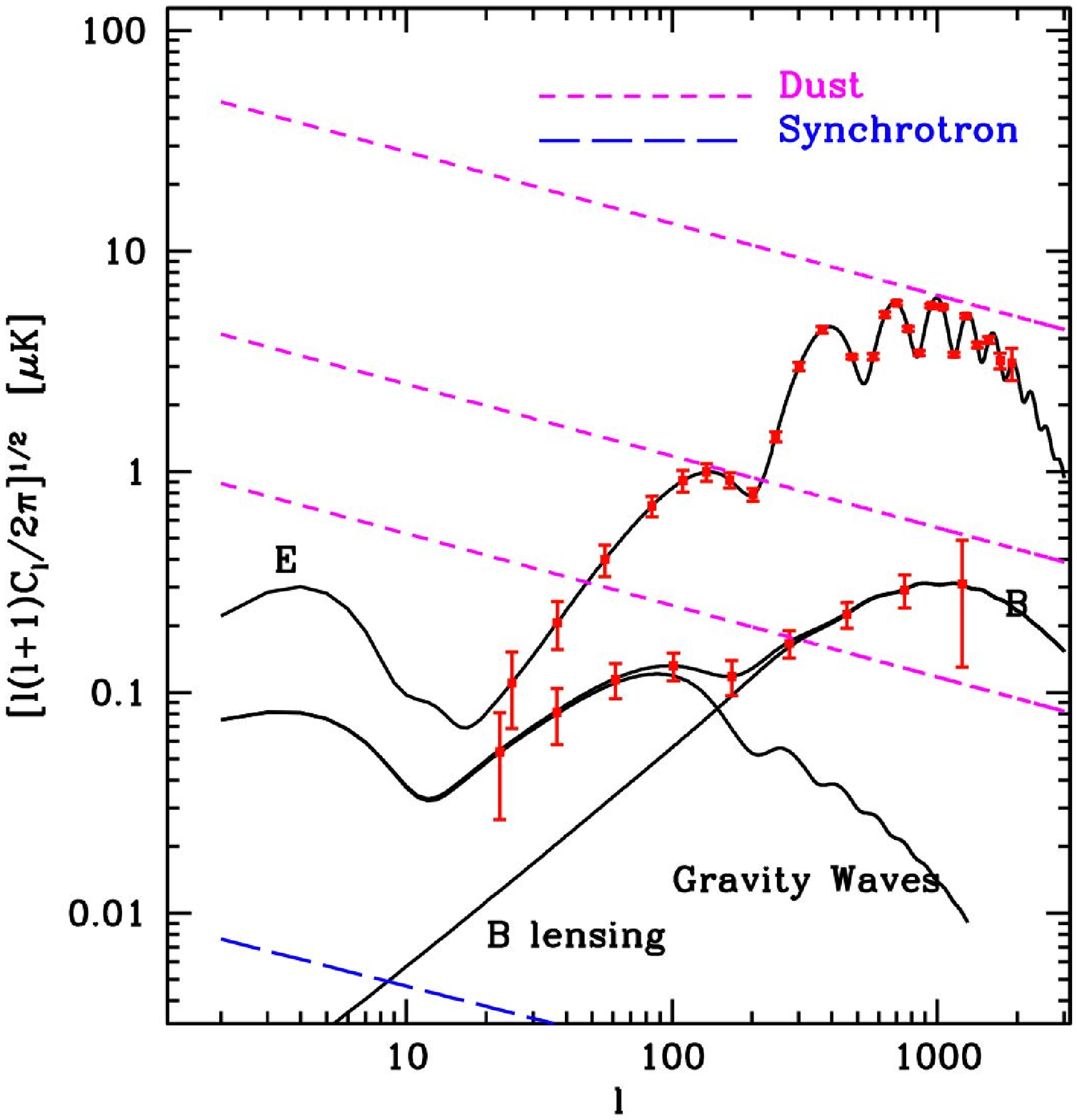}
}
\hspace{0cm}
\subfigure[] % caption for subfigure 
{
    \label{sci:sub:b}
    \includegraphics[height=2.2in]{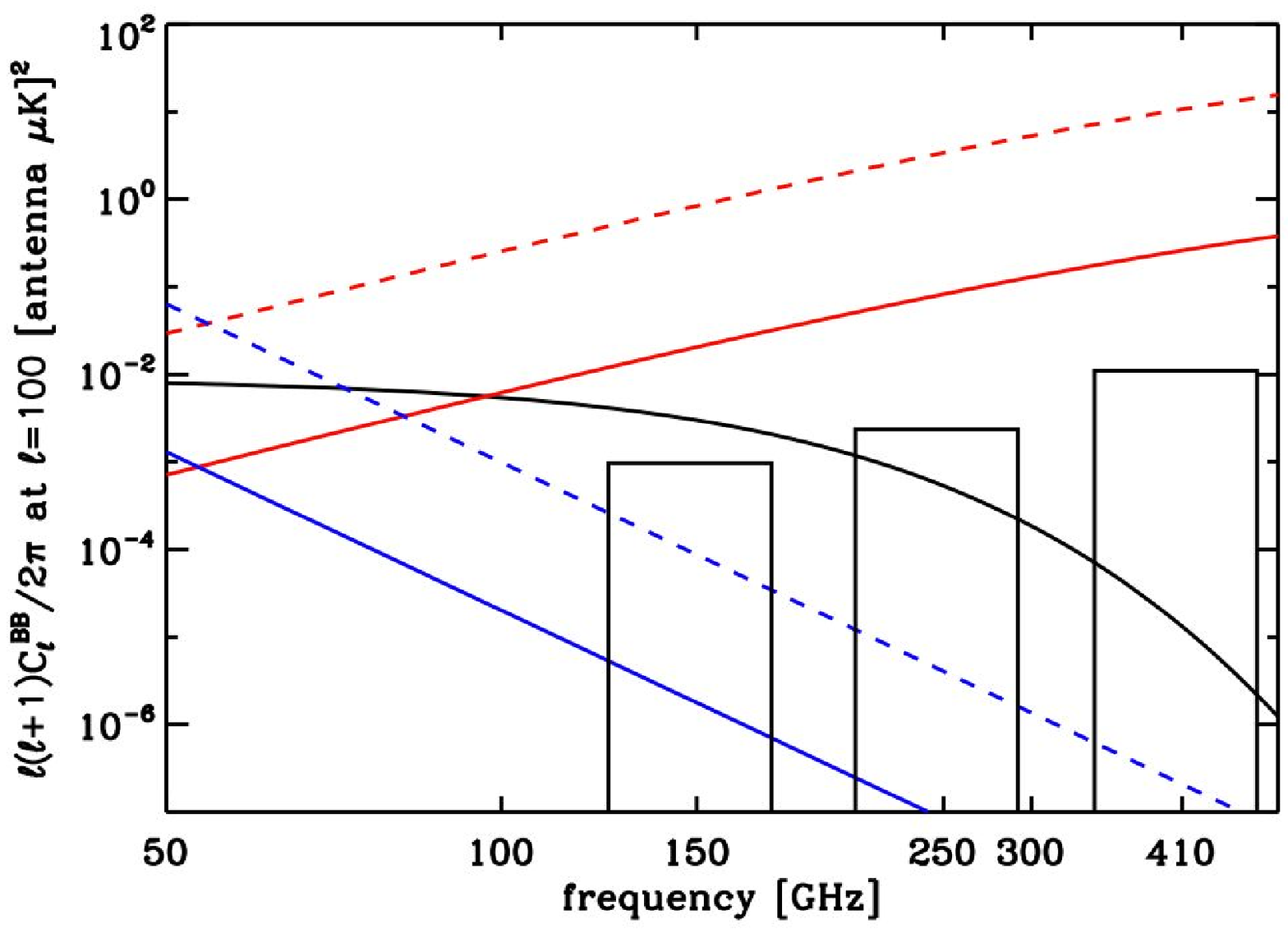}
}
\caption{\textit{a}:  Theoretical curves for $C_{EE}$ and $C_{BB}$ signals (assuming $r = 0.1$, defined at $\ell$=2) and predicted data points with error bars for a 14-day EBEX flight. The pink short dash and blue long dash lines show the anticipated $B$-mode power spectra of the synchrotron (at 150 GHz) and dust (at 150, 250 and 410 GHz, from bottom to top) foregrounds in the EBEX CMB patch. Foreground levels are extrapolated from the foreground fits of Gold et al.~\cite{Gold:2010}, scaled to the appropriate frequency and by the ratio of foreground intensity variance of the EBEX patch to the area outside WMAP's P06 mask.  As WMAP's detection of polarized dust emission has low significance~\cite{Gold:2010}, these levels are expected to be upper bounds.  The predicted error bars include contributions from instrument noise, sample variance and foreground subtraction (see text).  \textit{b}:  The anticipated $B$-mode power spectra of synchrotron (blue) and dust (pink) are shown for a large region of the sky outside WMAP's P06 mask (dotted) and in the EBEX CMB patch (solid) at a scale of $\ell$=100, where the gravitational wave $B$-mode signal is expected to peak.  Foreground levels are extrapolated from WMAP as described above.  The black curve shows the theoretical CMB $B$-mode signal for $r=0.1$. The bars show the expected EBEX instrument noise per band for a 14-day flight.} 
\end{figure}

In the inflationary paradigm, quantum fluctuations in the metric and the inflaton were stretched to astronomical sizes producing a spectrum of primordial fluctuations that can be decomposed into scalar, vector\footnote{Since vector perturbations are diluted by the expansion of space before the epoch of recombination and are not expected to be significant at the surface of last scattering in the simplest inflationary models, they will not be addressed.}, and tensor components.  The scalar component produced only $E$-mode polarization in the CMB.  However, the gravitational waves predicted by inflation, if they were present, produced both $E$- and $B$-modes in roughly equal strength~\cite{Kamionkowski:1997, Zaldarriaga:1997}, making the $B$-mode signal a unique probe of the inflationary epoch.  Figure \ref{sci:sub:a} shows theoretical power spectra for the $E$ and $B$ auto-correlations, $C_{EE}$ and $C_{BB}$, assuming a tensor to scalar ratio, $r$, of 0.1.  The $B$-mode curve is the sum of contributions from the primordial gravitational waves, peaking at about $\ell$ = 100, and the lensing $B$-mode signal, peaking at higher $\ell$.  Predicted data points with error bars are shown for EBEX for a 14-day flight.  

The amplitude of the gravitational wave component of $C_{BB}$ provides a direct measure of $r$.  Since different inflationary models and alternatives to inflation predict varying values of $r$, CMB polarization measurements can be used to distinguish between, and therefore allow or rule out, different classes of models.  Additionally, in the simplest inflationary models the value for $r$ provides the energy scale at which inflation occurred:  $V^{1/4}\propto~10^{16}~r^{1/4}$~GeV~\cite{Carroll:2004}.  If $r$ is very small, the standard deviation of the EBEX likelihood is forecasted to be $\sigma_r = 0.013$ without, and $\sigma_r = 0.02$ with foreground subtraction. For a likelihood centered around zero, 99\% of the area falls below $r=0.035$ and $r=0.056$, respectively. These estimates do not include systematic uncertainties. In Section~\ref{sec:foregrounds} we discuss foreground subtraction in more detail.

\subsection{Probe Large Scale Structure and Structure Formation}
\label{lensing}

The random deflections of the CMB photons caused by gravitational lensing of the CMB at late times resulted in mixing of power between primordial $E$- and $B$-modes~\cite{Zaldarriaga:1998}, generating a $B$-mode signal which peaks at small scales, shown in Figure \ref{sci:sub:a}.  The degree of lensing provides a measure of the integrated gravitational potential along the line of sight, probing the geometry of the Universe and the growth of structure.  

The amplitude of the lensing $B$-mode signal is predicted to within about 20\% based on current CMB measurements and other probes of large scale structure~\cite{Komatsu:2010, Tegmark:2004}.  EBEX will constrain the lensing $B$-mode amplitude within 6.5\% up to about $\ell$=900 if the amplitude of the signal is as expected, including anticipated uncertainties due to foreground subtraction.  The lensing $B$-mode measured by EBEX may constrain the dark energy equation of state~\cite{Acquaviva:2006}, and EBEX data used along with {\sl Planck}~\cite{Planck:2006} data is expected to improve constraints on the sum of the neutrino mass species, $\Sigma m_\nu$.  Additionally, since the shape and the amplitude of the lensing $B$-mode signal is well constrained by other data, a measurement of the lensing $B$-mode signal will provide a proof of concept of our ability to reconstruct $E$- and $B$-modes on a partial patch of the sky in the presence of noise, instrumental effects, and foregrounds.

\subsection{Characterize Polarized Emission from Dust in our Galaxy}

Although polarized galactic foreground sources at sub-millimeter wavelengths have been measured by a number of experiments, current knowledge of the frequency, scale, and spatial dependence of foregrounds at the EBEX observing frequencies is limited.  During the long duration science flight EBEX will map a patch of sky with an exceptionally low expected mean brightness, on the order of 3 $\mu K$.  Figure \ref{sci:sub:a} shows the anticipated scale dependence of the $B$-mode power spectra of the synchrotron (at 150 GHz) and dust foregrounds (at 150, 250 and 410 GHz) in the EBEX CMB patch.  The anticipated frequency scaling of the foreground $B$-mode signal and the significantly lower amplitude of emission in the EBEX patch are shown in Figure \ref{sci:sub:b}.  The plot shows that over the EBEX observing bands the dust foreground dominates the synchrotron, which is expected to be negligible, allowing for a high signal to noise detection of the polarized dust signal.   

\section{INSTRUMENT}

%\subsection{EBEX Experimental Approach}

\begin{figure}[t!]
\centering
\subfigure[] % caption for subfigure a
{
    \label{gond:sub:a}
    \includegraphics[height=3.0in]{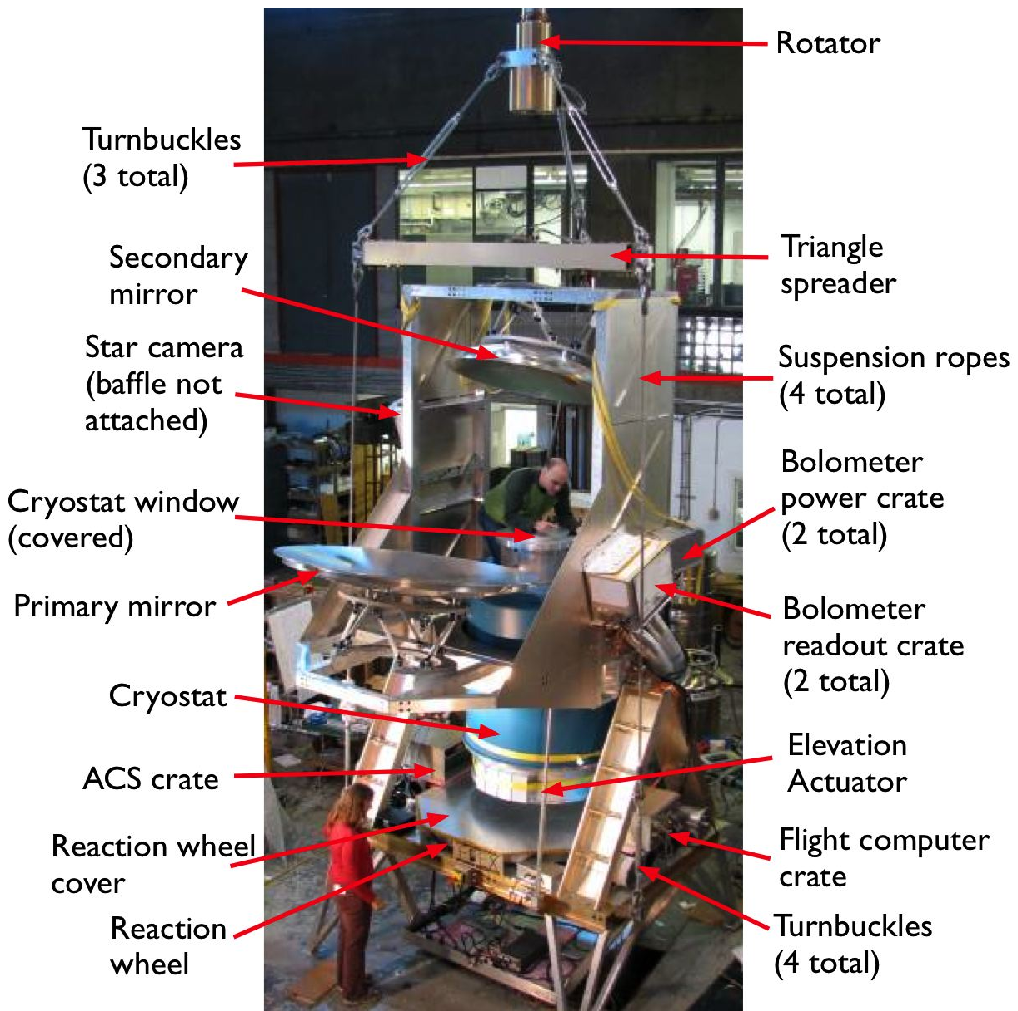}
}
\hspace{0cm}
\subfigure[] % caption for subfigure b
{
    \label{gond:sub:b}
    \includegraphics[height=2.6in]{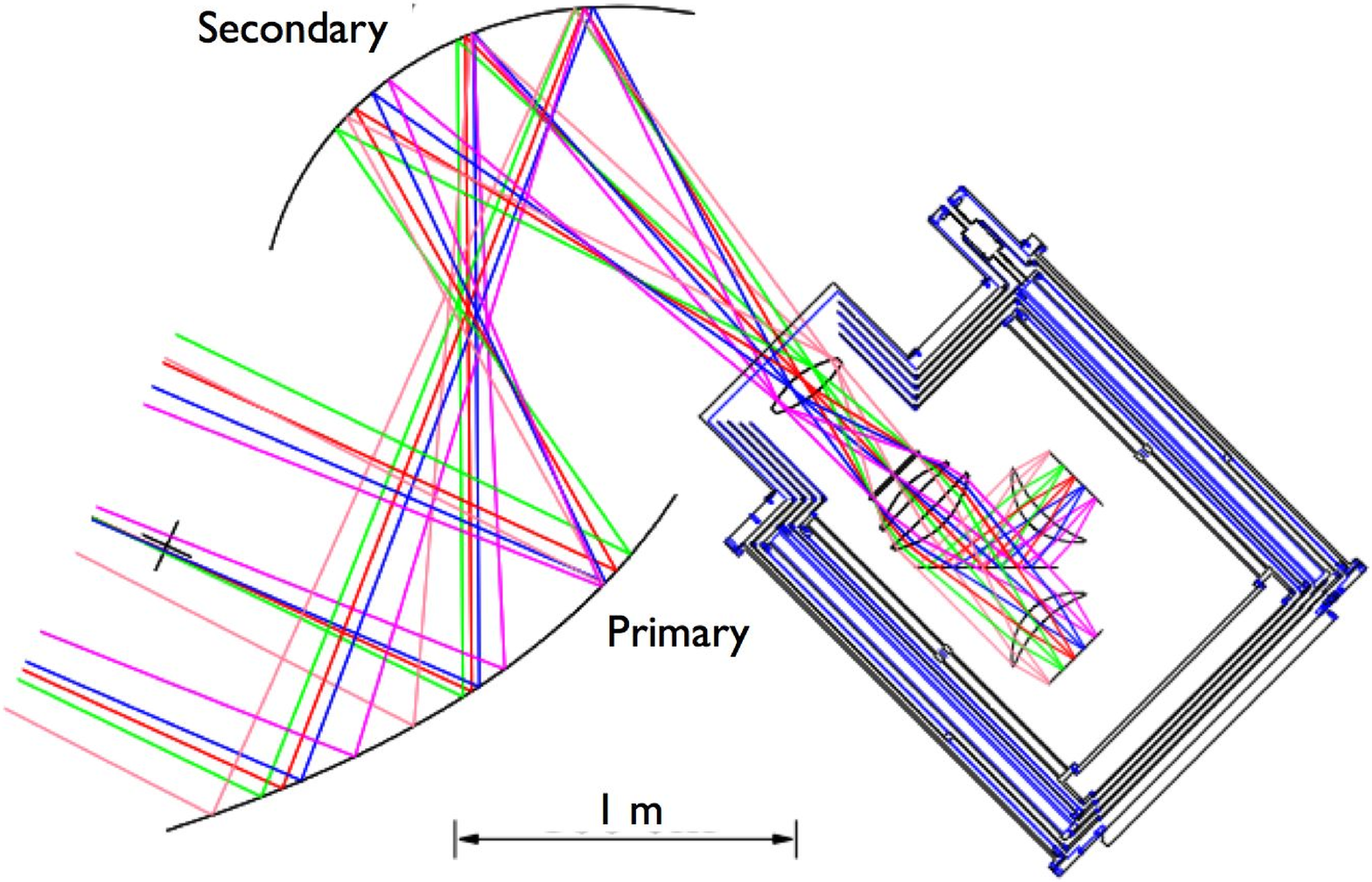}

}
\caption{{a:}  The EBEX gondola during system integration.  \textit{b:}  A ray diagram of the EBEX warm and cold optics.  The primary and secondary mirror are at ambient temperature and all the optics inside of the cryostat are at temperatures below 4 K.}
\label{gondolalabels}
\end{figure}

The design of the EBEX instrument, shown in Figure \ref{gond:sub:a}, and scan strategy were driven by the need for a high sensitivity microwave polarimeter with a low susceptibility to systematic effects and a sensitivity to a wide range of scales on the sky.  A summary of the EBEX instrument and flight parameters is provided in Table \ref{ebexsummary}.  The $\sim$~2,700 kg (6,000 lb) instrument will map the polarization of the CMB while suspended from a 1 Mm$^3$ (34 Mft$^3$) Helium balloon at the top of the stratosphere $\sim$35-40 km over Antarctica.  Observing the microwave sky from this altitude allows for observation at high frequencies; this means EBEX will be sensitive to only one significant foreground--dust, assuming recent foreground measurements from {\sl WMAP}~\cite{Gold:2010}.  Additionally, observing during the polar summer over Antarctica provides relatively constant temperatures throughout the 24 hour diurnal cycle, allowing the balloon to maintain an approximately constant altitude, and thus constant atmospheric loading on the detectors.  
\begin{table}[t!]
\begin{center}
\vspace{0.1in}
\caption{Anticipated instrument characteristics and flight properties during the long duration flight.} 
\vspace{0.1in}
\begin{tabular}{p{2.8in}p{0.6in}p{0.6in}p{0.6in}}
\hline
%\multicolumn{4}{c}{\textbf{Anticipated EBEX Parameters During the Long Duration Flight}}\\
%\hline
{Nominal Band Frequencies (GHz)}&\multicolumn{1}{c}{150}&\multicolumn{1}{c}{250}&\multicolumn{1}{c}{410}\\
{Number of Detectors (Light)$^a$} &\multicolumn{1}{c}{768}&\multicolumn{1}{c}{384}&\multicolumn{1}{c}{280}\\
{Beam Size (arcmin)} &\multicolumn{3}{c}{8}\\
{Error per beam size pixel$^{b}$ ($\mu$K) Q/U}, T&\multicolumn{3}{c}{1.3, 0.9}\\
%{NE$Q,U$/detector$^b$ ($\mu K\sqrt{s}$)}&\multicolumn{1}{c}{136}&\multicolumn{1}{c}{282}&\multicolumn{1}{c}{2180}\\
%{NET/detector$^b$ ($\mu K\sqrt{s}$)} &\multicolumn{1}{c}{96}&\multicolumn{1}{c}{199}&\multicolumn{1}{c}{1538}\\
{Total Sky Coverage (deg$^2$)}&\multicolumn{3}{c}{$\sim$1\% of the sky} \\
{Flight Duration at Float$^c$ (days)}&\multicolumn{3}{c}{14}\\
\hline
\label{ebexsummary}
\end{tabular}
\end{center} 
$^a$   Number of light detectors refers to the total number of detectors exposed to radiation through the cryostat window and read out using a multiplexing factor of 12.\\
$^b$  Includes detectors with a Strehl ratio of 0.85 or above and based on a 14-day long duration flight.\\
$^c$  Does not include loss of time due to cycling of refrigerators and tuning of bolometers and SQuIDS.  Refrigerator cycling, which takes approximately 4 hours, is expected to occur twice during the flight.
\end{table}

\subsection{Warm Optics}
The EBEX telescope is a compact off-axis Gregorian Mizuguchi-Dragone system, chosen to minimize polarized systematic effects~\cite{Hanany+Marrone:2002, Tran:2003}.  The telescope has a 6$^{\circ}$ field of view with high image quality, as defined by the Strehl ratios of 0.85 or above for the 1432 detectors across the large focal plane.  The 1.5 m primary mirror allows for sensitivity to both the gravitational wave and lensing $B$-mode signals.  Figure \ref{gond:sub:b} shows a ray diagram of all of the EBEX optics, including the warm mirrors and the cold optics in the cryostat.  The telescope mirrors are aligned and supported by adjustable hexapods.    
%Alignment of the warm optics is achieved using inside micrometer measurements between tooling balls on the mirrors and adjustments of hexapods, which also provide mechanical support. 
%\begin{figure}[h!]
%\begin{center}
%\includegraphics[height=2.8in]{figureswarmoptics/warmoptics.jpg}
%\caption[Ray diagram of the EBEX optics]{An optical simulation ray diagram showing all of the EBEX optics, including the warm mirrors and the cold optics in the cryostat.}
%\label{warmoptics}
%\end{center}
%\end{figure}

\subsection  {Cryostat and Cold Optics}
\label{coldoptics}
\label{focalplane} 

\subsubsection{Overview} 

\begin{figure}
\centering
\subfigure[ ] % caption for subfigure a
{
    \label{cryo:sub:a}
     \includegraphics[height=2.7in]{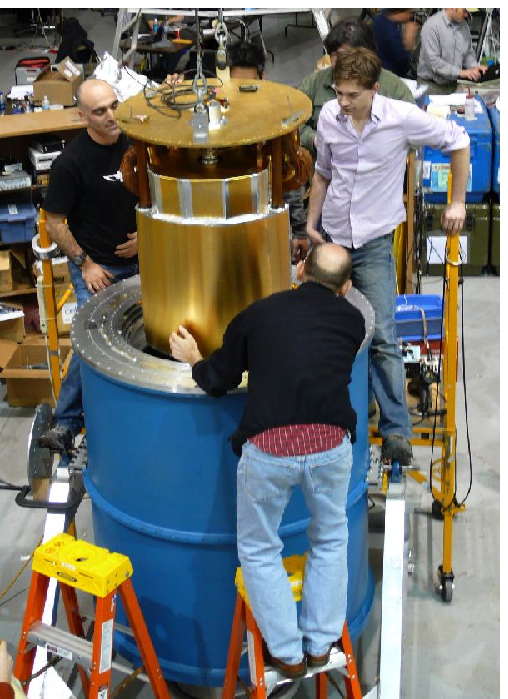}
}
\hspace{0cm}
\subfigure[] % caption for subfigure b
{
    \label{cryo:sub:b}
    \includegraphics[height=2.7in]{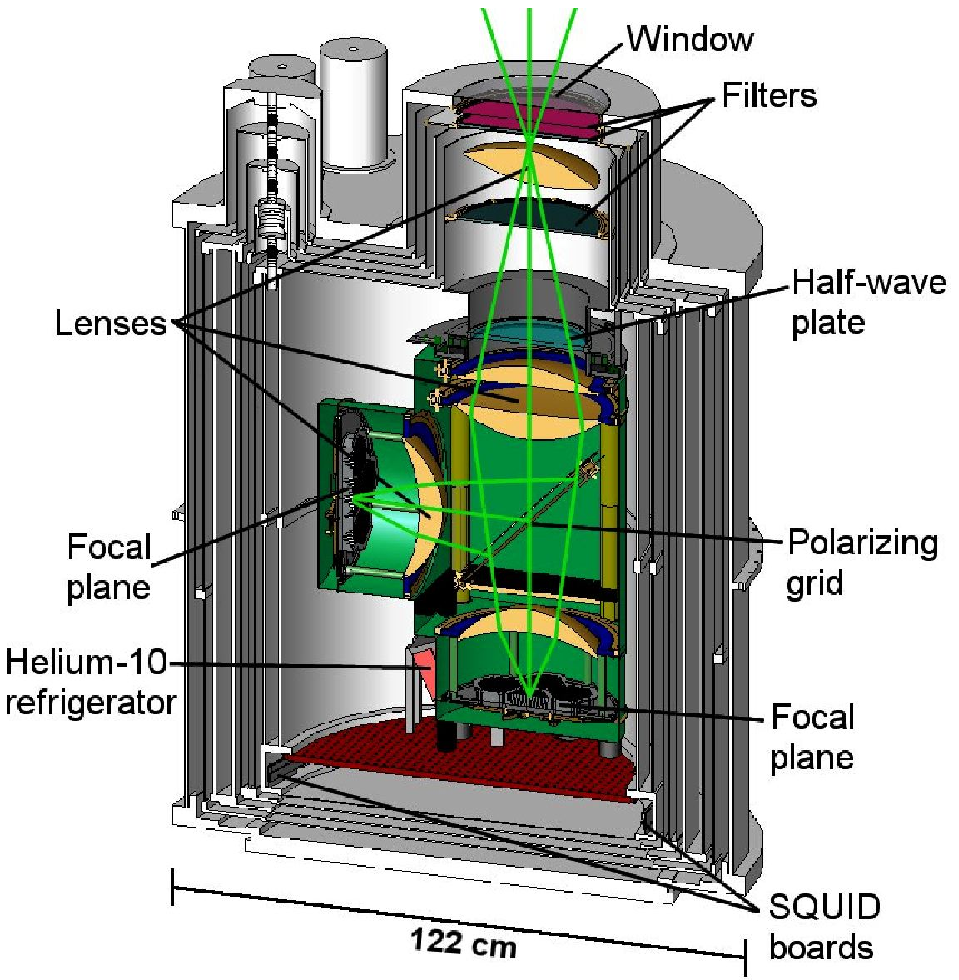}
}
\caption{\textit{a:}  Installation of the EBEX receiver in the cryostat during integration in Ft. Sumner, NM.  \textit{b:}  Cutaway drawing showing the cryostat and receiver.}
\end{figure}
\begin{figure}[h!]
\centering
\subfigure[ ] % caption for subfigure a
{
    \label{opticsetc:sub:a}
     \includegraphics[height=2.6in]{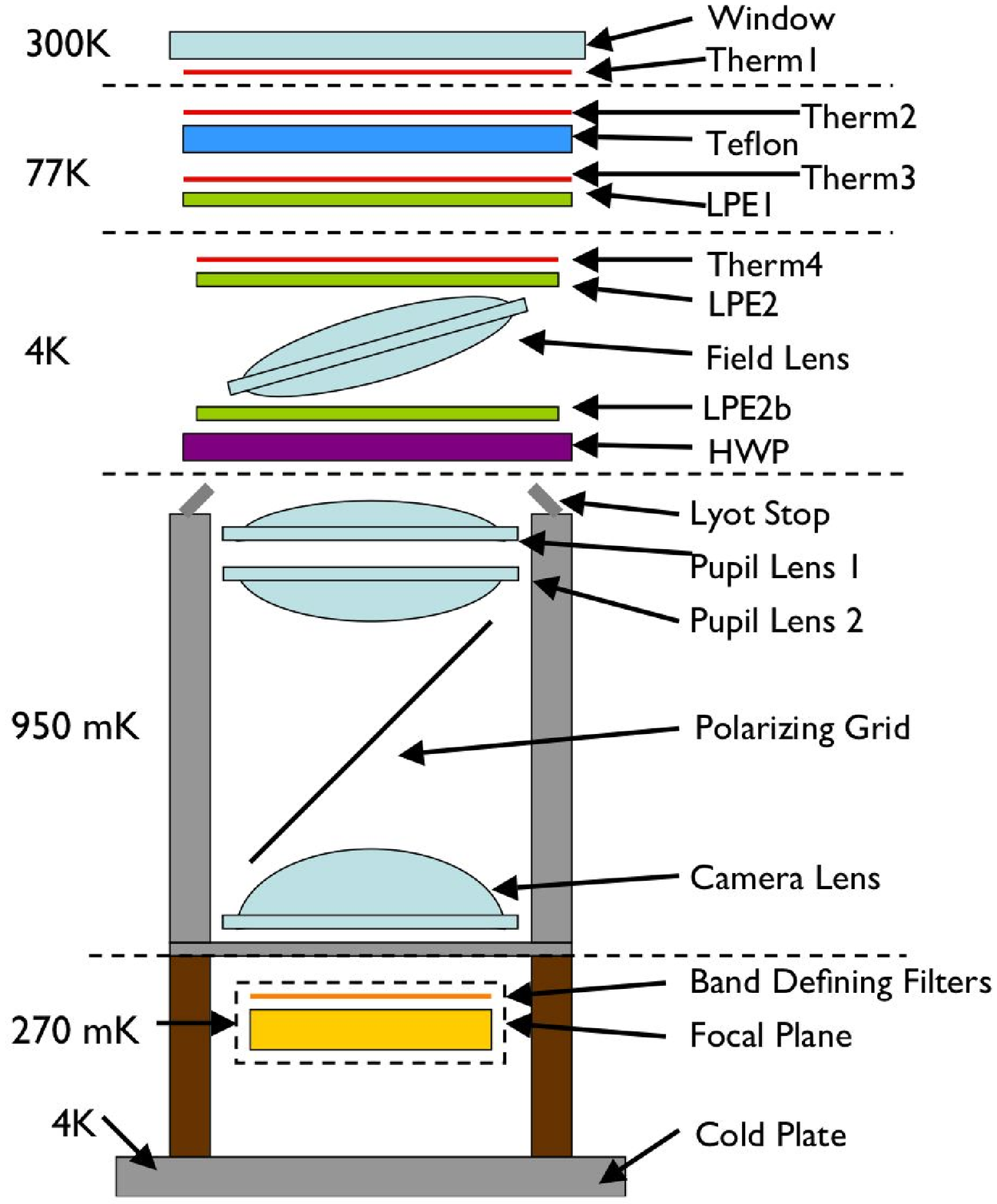}
}
\hspace{0cm}
\subfigure[] % caption for subfigure b
{
    \label{opticsetc:sub:b}
    \includegraphics[height=2.6in]{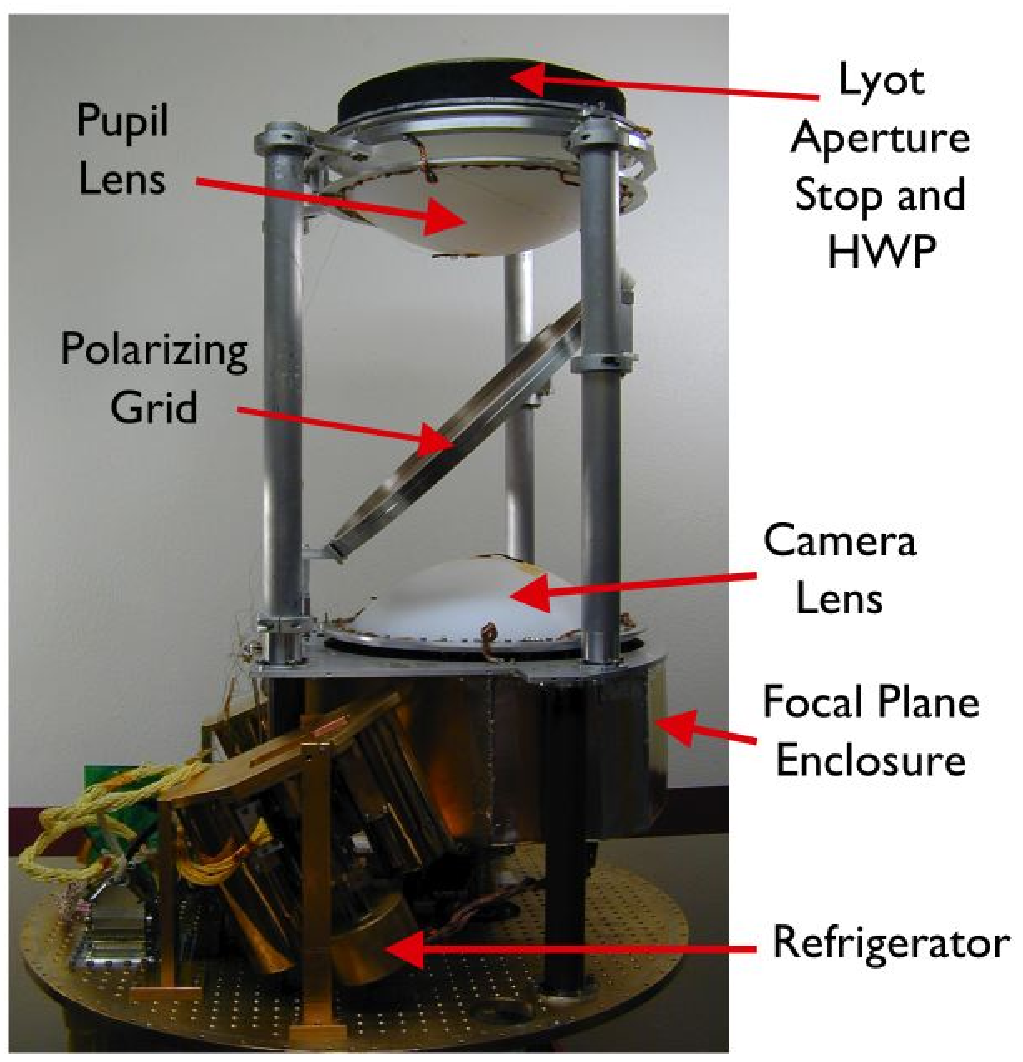}
}
\caption[The filtering scheme and optics box]{\textit{a:}   Drawing of the EBEX thermal (Therm) and low-pass edge (LPE) filters and lenses in the engineering flight configuration.  \textit{b:}  The EBEX optics box shown in the engineering flight configuration with one of the two focal planes that will be present in the long duration flight.} 
\end{figure}
The EBEX receiver, shown during installation in Figure 3(a), is housed in a liquid nitrogen and liquid helium cryostat; Figure \ref{cryo:sub:b} shows a cutaway view of the cryostat.  CMB photons enter the cryostat through a 30~cm diameter window made from ultra high molecular weight polyethylene (UHMWPE).  During the engineering flight a single thick window was used.  However, for the long duration flight a double window mechanism will be implemented allowing for use of a thinner window during observations at float altitudes to reduce the absorption of sky signal and thermal loading of the window on the detectors.  Below the window, a stack of thermal (labeled Therm1 through Therm4) and low-pass edge (labeled LPE1 to LPE2b) metal mesh filters are distributed amongst the 300, 77, and 4 K cryogenic stages, shown in the engineering flight configuration in Figure \ref{opticsetc:sub:a}, to minimize the thermal load on the cryogens. 

\begin{figure}[t!]
\centering
\subfigure[ ] % caption for subfigure a
{
    \label{focalplane:sub:a}
     \includegraphics[height=2in]{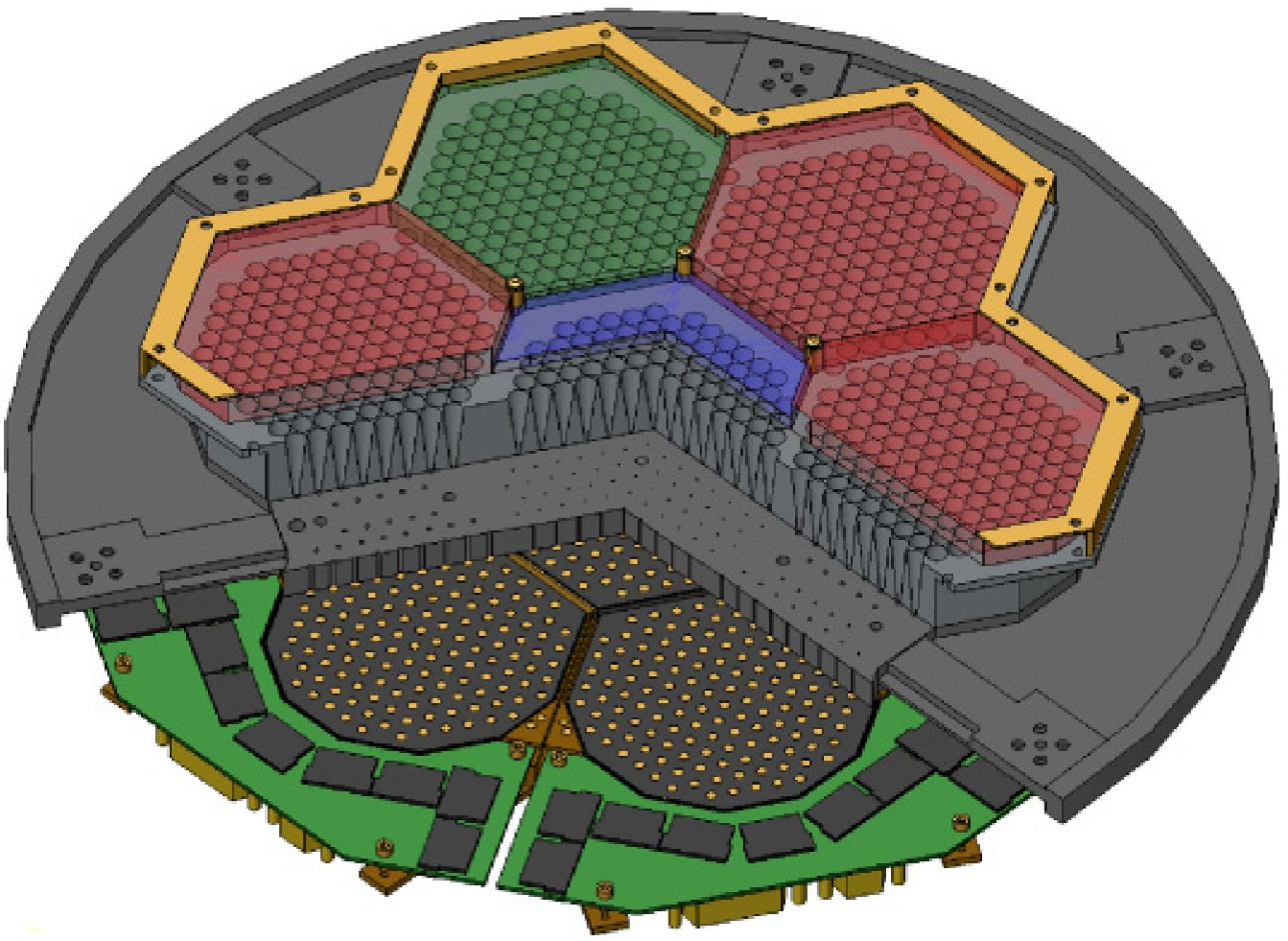}
}
\hspace{0cm}
\subfigure[] % caption for subfigure b
{
    \label{focalplane:sub:b}
    \includegraphics[height=2in]{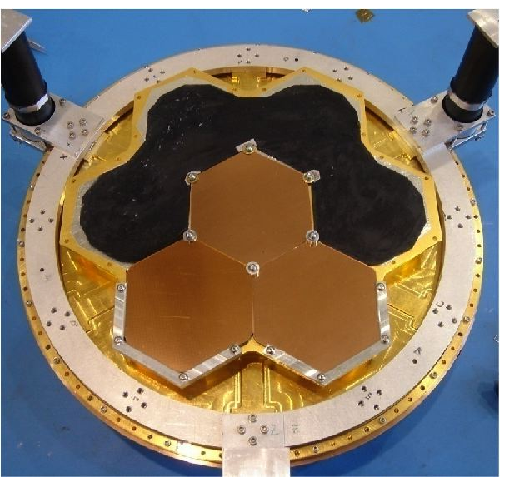}
}
\caption[The focal plane]{\textit{a:}  Three dimensional CAD drawing of the EBEX focal plane.  The colors encode the frequency of the band defining filters; red is 150 GHz, green is 250 GHz, and blue is 410 GHz.  \textit{b:}  The partially populated EBEX focal plane in the engineering flight configuration with three TES wafers and band defining metal mesh filters installed.}\label{focalplanephoto}
\end{figure}

A cold Lyot stop at 4 K is located at the top of the optics box, shown in Figure \ref{opticsetc:sub:b}, to minimize the presence of sidelobes in the beam.  At the stop a continuously rotating half-wave plate (HWP) modulates the polarization.  Next, a polarizing grid oriented at 45$^{\circ}$ to the incoming beam splits the radiation into horizontally and vertically polarized components before the radiation is incident on one of the two identical focal planes; Figures \ref{opticsetc:sub:a} and \ref{opticsetc:sub:b} show the receiver in the engineering flight configuration with one focal plane installed.  Metal mesh band-defining filters at 150, 250 and 410 GHz are located on the top of the focal planes, shown in Figures \ref{focalplane:sub:a} and \ref{focalplane:sub:b}; the width of each band is about 25\%.  Below the filters the radiation couples to the bolometric detector arrays through smooth-walled conical feedhorns and cylindrical wave guides; the feedhorns fill a smaller fraction of the primary mirror at higher frequencies, producing an 8' beam at all frequencies.  Re-imaging lenses made out of UHMWPE are mounted above the cold stop and in the optics box to produce an image on the flat focal plane. 

Temperature stages at 240 K, 77 K, 30 K and 4 K are produced by the liquid nitrogen and helium cryogens and by passing their associated boil-off gas through vapor cooled shields.  A two-stage adsorption refrigerator that consists of a $^4$He 
refrigerator backed by a second $^4$He refrigerator cools the the optics box, shown in Figure~\ref{opticsetc:sub:b}, to $\sim$1~K to reduce thermal loading on the focal plane.  A three-stage adsorption refrigerator that consists of a $^3$He stage backed by $^3$He and $^4$He refrigerators keeps the focal plane at 300 mK.  The cryogenic performance and housekeeping electronics are discussed in detail elsewhere~\cite{Sagiv:2010}.

The cold optics alignment is verified using coordinate-measuring machine (CMM) measurements of the relative positions of the focal plane and lenses to reference points external to the cryostat, which allows for indexing of the cold optics to the warm optics.  Before the engineering flight, CMM measurements verified that the optical alignment of all cold optics components was within 0.127 mm (0.005 in) in translation and 0.1$^{\circ}$ in rotation of the nominal positions in the optics model.  The cold and warm optics are referenced via measurements to tooling balls mounted to the top of the cryostat.

\subsubsection  {Polarimetry}
\label{sec:hwp}

\begin{figure}[t!]
\centering
\subfigure[ ] % caption for subfigure a
{
    \label{hwp:sub:a}
     \includegraphics[height=2.1in]{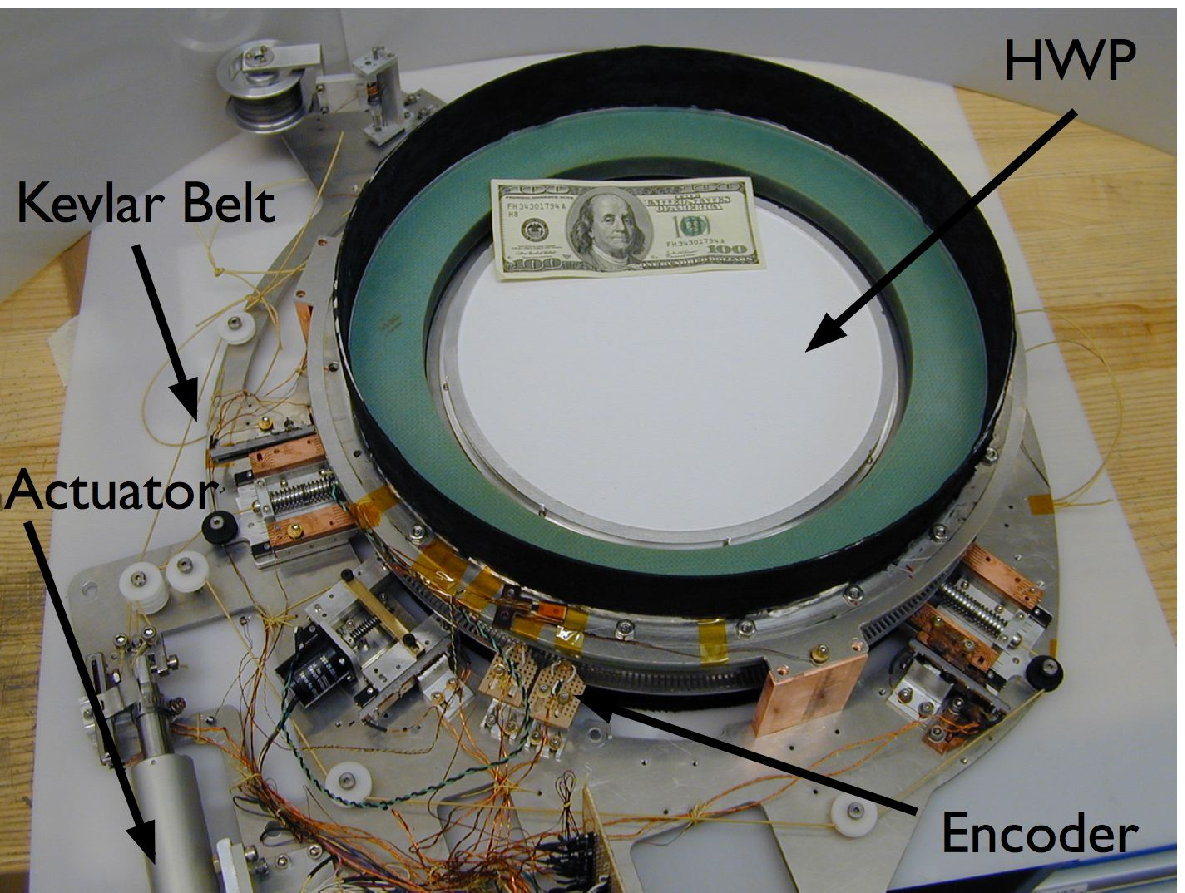}
}
\hspace{0cm}
\subfigure[] % caption for subfigure b
{
    \label{hwp:sub:b}
    \includegraphics[height=2.1in]{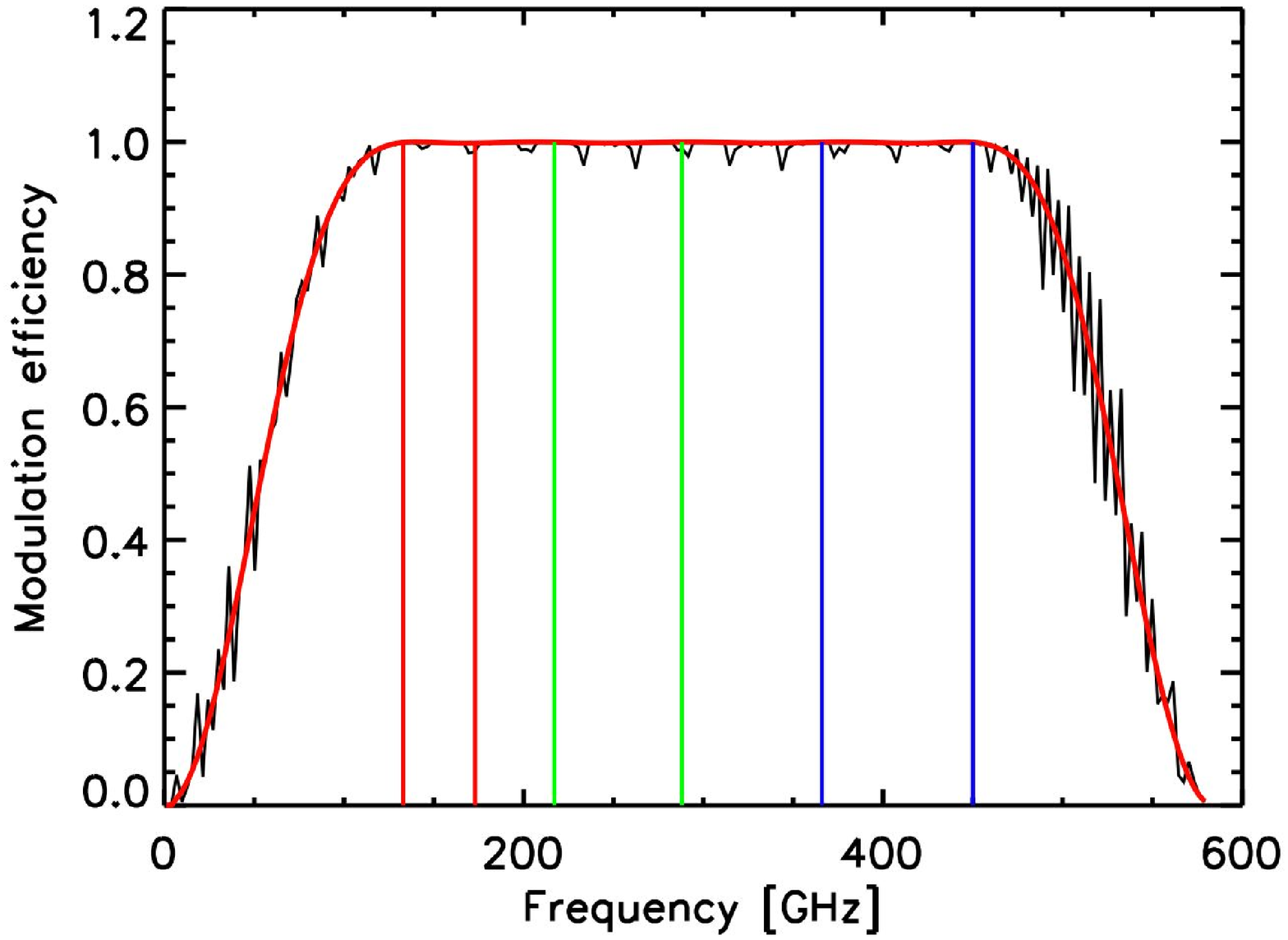}
}
\caption{\textit{a:}  The EBEX HWP.  \textit{b:}  Predicted modulation efficiency over the EBEX bands for a five-stack HWP with a single layer thickness of 1.62 mm.  The black curve includes reflections between the five layers in the stack while the red does not.}
\end{figure}

\begin{figure}[t!]
\centering
\subfigure[ ] % caption for subfigure a
{
    \label{hwp2:sub:a}
     \includegraphics[height=1.9in]{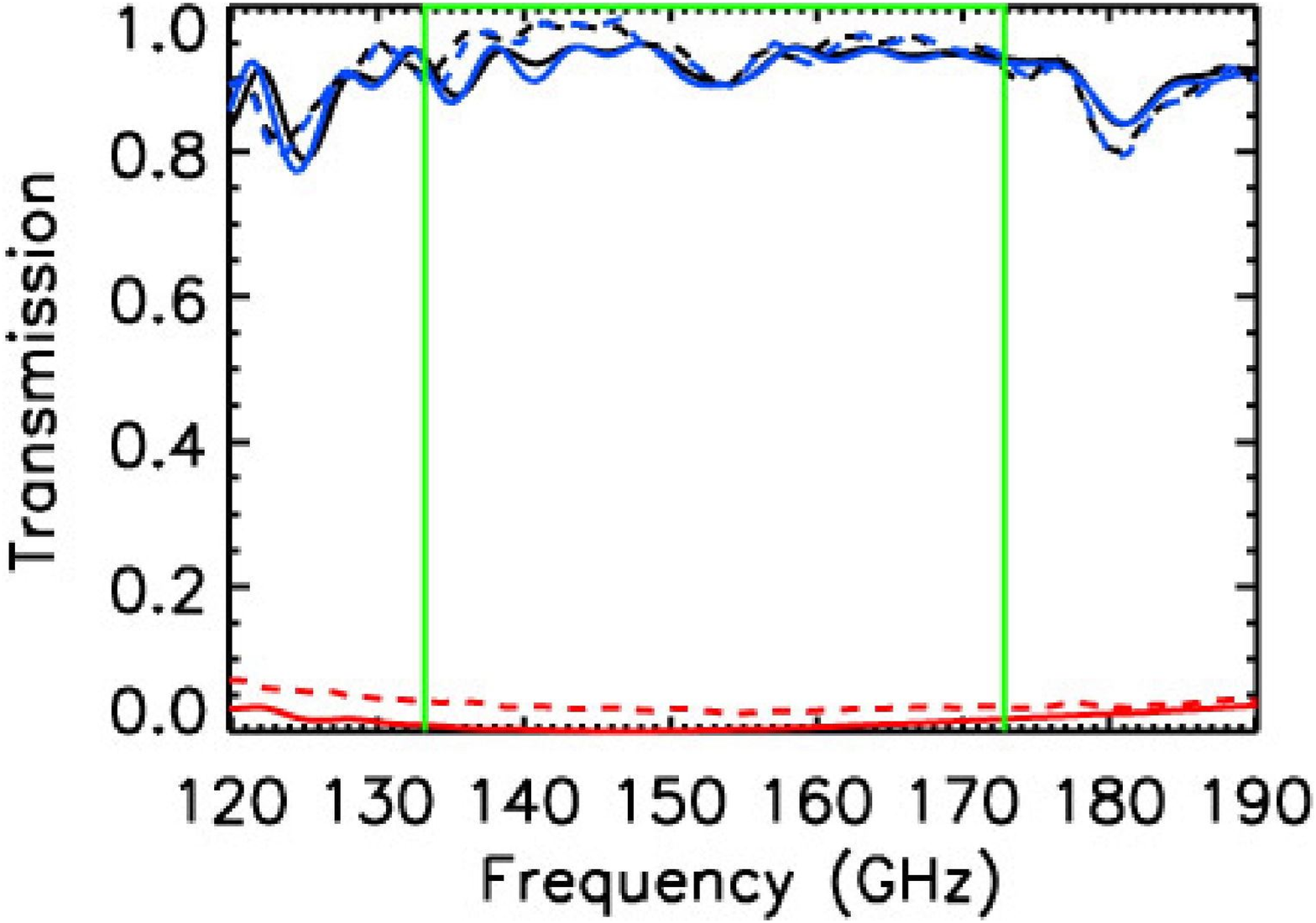}
}
\hspace{0cm}
\subfigure[] % caption for subfigure b
{
    \label{hwp2:sub:b}
    \includegraphics[height=2.6in]{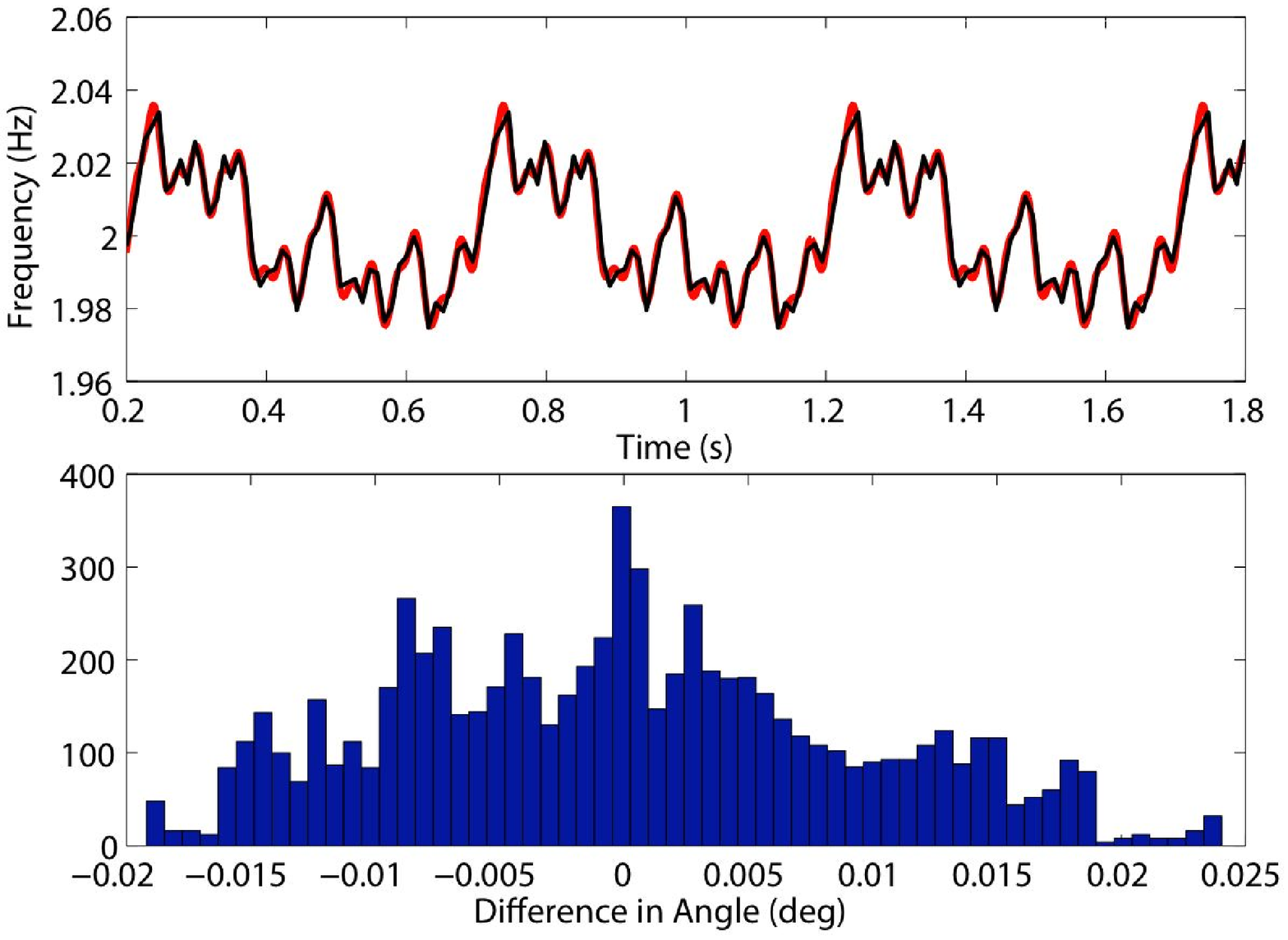}
}
\caption{\textit{a:}  Transmission spectra of measurements (dashed) and theoretical model fits (solid) for the EBEX HWP at 300 K for input polarizations parallel to (black), at 45$^{\circ}$ to (red), and perpendicular to (blue) the HWP active axis over the 150 GHz band~\cite{Matsumura:2009}.  \textit{b:}  Simulation of HWP rotation and angle reconstruction. \textit{Top} Time stream of the input frequency of HWP rotation (red) and the recovered rotation frequency extracted from decoding the rotation angle (black). \textit{Bottom} Histogram of the difference between the input and recovered HWP angle; the standard deviation is 0.03$^{\circ}$. The input data to the simulation had the same noise level and an order of magnitude higher HWP rotation speed variation compared with the data taken during the engineering flight.}
\end{figure}

To mitigate systematic effects and to reduce noise the polarized CMB signal is modulated using a HWP and analyzed using a wire grid polarizer; this is discussed in more detail in Sec. \ref{systematics}.  The EBEX HWP, shown in Figure~\ref{hwp:sub:a}, is made broadband using five layers of sapphire crystal with active axes rotated\footnote{The relative rotation axes from one surface to another are 0$^{\circ}$, 25$^{\circ}$, 88$^{\circ}$, 25$^{\circ}$, and 0$^{\circ}$.} with respect to one another and bonded using polypropylene.  The resulting predicted modulation efficiency\footnote{The efficiency of the HWP is defined as the ratio of the difference over the sum of parallel and perpendicular temperature.} of the EBEX achromatic HWP is 98\% from 120 to 450 GHz; Figure \ref{hwp:sub:b} shows the theoretical modulation efficiency over the EBEX observing bands.  Figure \ref{hwp2:sub:a} shows a measurement at 300K of the transmission through the EBEX HWP across the 150 GHz frequency band for input polarizations at 0$^{\circ}$, 45$^{\circ}$, and 90$^{\circ}$ to the HWP active axis.  The figure also includes the results of a fit of the data to a theoretical model which includes the thickness and angular orientation of the plates in the 5-stack and the thickness of the anti-reflection (AR) coatings.  The calculated efficiency of the HWP at 4 K based on the fit of the data taken at 300K to the theoretical model is 1, 1, and 0.98 at 150, 250, and 410 GHz, respectively~\cite{Matsumura:2009}.


The HWP is placed at the 4 K Lyot stop to minimize polarized systematics.  A superconducting magnetic bearing (SMB) consisting of a magnetic ring and a high temperature superconductor~\cite{Hanany:2003} supports the HWP and allows for continuous rotation at 2 Hz.  A motor outside of the cryostat drives a shaft that penetrates the cryostat and turns the rotor of the SMB via a tensioned kevlar belt.  An encoder records the angular position of the HWP.  Figure \ref{hwp2:sub:b} shows the results from a simulation of HWP angle reconstruction where the input data had noise and speed variation similar to the data taken during the engineering flight. We find that our method of HWP position encoding and subsequent angle reconstruction allows for HWP position reconstruction to an accuracy better than 0.03$^{\circ}$.  Demodulation of a polarized signal using a reconstructed HWP timestream has been successfully demonstrated by the MAXIPOL team with their flight data~\cite{Johnson:2007}. 

\subsection{Detectors and Read Out Electronics}

\subsubsection  {Bolometric Detector Arrays}

\begin{figure}[t!]
\begin{center}
\includegraphics[height=1.9in]{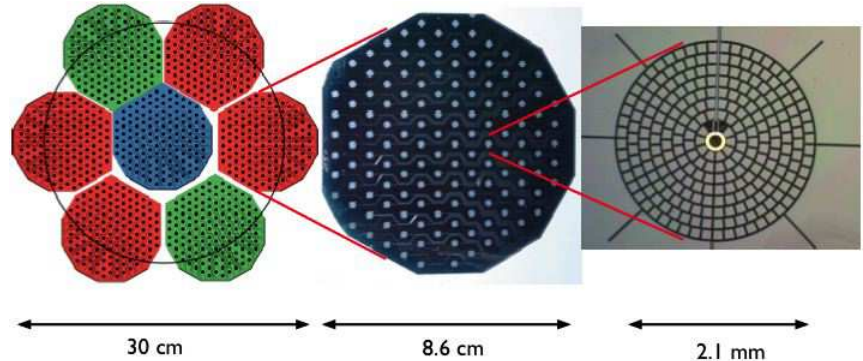}
\caption[The detector wafers]{\textit{Left}:  A drawing of the EBEX detector wafers on one of the two identical focal planes, with color coding to indicate the frequency of the band-defining filter above the wafer; red is 150 GHz, green is 250 GHz, and blue is 410 GHz.  A Strehl ratio of 0.85 or above is achieved at detectors within the black circle.  \textit{Middle}:  A decagon detector wafer containing 140 TES bolometers.  \textit{Right}:  A single detector.  (Figure courtesy Clayton Hogen-Chin).} \label{fpdetector}
\end{center}
\end{figure}

The TES bolometric detectors are tightly packed into arrays.  Figure \ref{fpdetector} shows a drawing of seven of the detector array wafers assembled on one of the two identical focal planes.  The wafers, designed and fabricated at the University of California at Berkeley, are produced using thin film deposition and optical lithography on silicon.  Table \ref{ebexsummary} summarizes the number of detectors per band and the sensitivity per pixel for the long duration flight. 

An EBEX transition edge sensor (TES) bolometer is shown in the right panel of Figure \ref{fpdetector}.  The metalized silicon nitride spider web absorber is designed for low heat capacity, and thus a fast optical time constant, and reduced susceptibility to cosmic ray hits.  Silicon nitride legs provide the thermal link between the absorber and the focal plane heat sink.  The aluminum/titanium TES is connected to superconducting leads that extend to the wafer periphery and to the gold ring which is visible in the middle of the detector; the gold ring is used to increase the sensor heat capacity to provide stability.  A more detailed discussion of the working principles of TES bolometers and the EBEX detector arrays is provided elsewhere~\cite{Hubmayr:2008, Aubin:2010}.

\subsubsection{Bolometer Readout Electronics}

The signals from many bolometers are digital frequency multiplexed (DfMUX) to reduce the total number of wires into the cryostat, thus limiting the heat load on the cryogens and minimizing power consumption.  A detailed overview of the design and performance of the DfMUX boards and related electronics is provided elsewhere~\cite{Hubmayr:2008, Aubin:2010}.  The readout electronics, designed at McGill University, provide a constant AC bias voltage at a unique frequency to each bolometer.  Microwave power incident on the detectors increases the temperature, and therefore resistance, of the TES, altering the current through the sensor.  The currents through a group of bolometers are measured by a superconducting quantum interference device (SQuID) amplifier array and the signals are digitized and demodulated.  During the engineering flight the bolometers were multiplexed in groups of 8, and a multiplexing factor of 12 is planned for the long duration flight.  Since the DfMUX boards dissipate a significant amount of power each board is well heat sunk to a plate at the back the readout crate.  Heat is conducted away from the crate to a plate that radiates to the sky; for the long duration flight we plan to circulate a liquid coolant.  

\subsection{Gondola and Attitude Control}

The EBEX gondola, made primarily of aluminum, was designed and fabricated at Space Sciences Lab (SSL) at the University of California at Berkeley.  The cryostat, warm optics and critical pointing sensors are mounted to a rigid inner frame tower which pivots on two trunnion legs mounted to an outer frame table, shown in Figure~\ref{gond:sub:a}.  The outer frame is suspended below a triangular spreader bar by lightweight polyethylene ropes, and the spreader is connected to the rotator by turnbuckles.  Motors in the rotator and reaction wheel, shown in the figure, drive sinusoidal azimuth scans at nearly constant speed, and a motor driven actuator moves the elevation of the inner frame from 15$^\circ$ to 68$^\circ$.  All electronics on the gondola will be powered by a solar power system during the long duration flight.  A detailed overview of the instrument control, data logging and telemetry software and hardware is provided elsewhere in these proceedings~\cite{Milligan:2010}.  

The attitude control system (ACS) provides control of the gondola and telescope boresight pointing data.  The ACS sensors planned for the long duration flight include a differential GPS system, a three-axis magnetometer, a sun sensor, a pair of redundant star cameras and two redundant sets of three orthogonally positioned fiberoptic rate gyroscopes.  The gyroscope rate data is used in real-time and in post-flight pointing reconstruction to interpolate between absolute sensor readings and for feedback in the azimuth control loop.  The real-time pointing accuracy exceeds the 0.5$^{\circ}$ requirement.  The predicted reconstruction pointing solution, achieved through post-flight analysis using a state model approach with the star cameras and gyroscopes dominating the solution, will exceed the 9" requirement by roughly an order of magnitude.
%\footnote{ Previous projects, such as BLAST, have had success using a Kalman filter approach~\cite{Pascale:2007}.  We are currently investigating the possibility of using a particle filter.} 
This 9" pointing requirement ensures that the systematic effects caused by pointing errors, which produces mixing of $E$ and $B$, are minimized.  The pointing sensors are indexed to the microwave beam using simultaneous microwave observations of calibrator sources and star camera solutions.  The gondola design and the ACS are discussed in detail elsewhere~\cite{RK:2010}.  

%Real-time azimuthal control is achieved using motors in the rotator and the reaction wheel.  The reaction wheel motor provides fine-tuned control while the rotator motor provides bursts of torque when large angular accelerations are required, such as at scan turnarounds, to prevent saturation of the reaction wheel.  The real-time elevation is controlled by a simple P feedback loop with an acceleration limit

\section{EBEX SKY COVERAGE, SCAN STRATEGY, and CALIBRATION SOURCES}
\label{scanstrategy}

\begin{figure}
\centering
\subfigure[] % caption for subfigure a
{
    \label{skycov:sub:a}
    \includegraphics[height=1.9in]{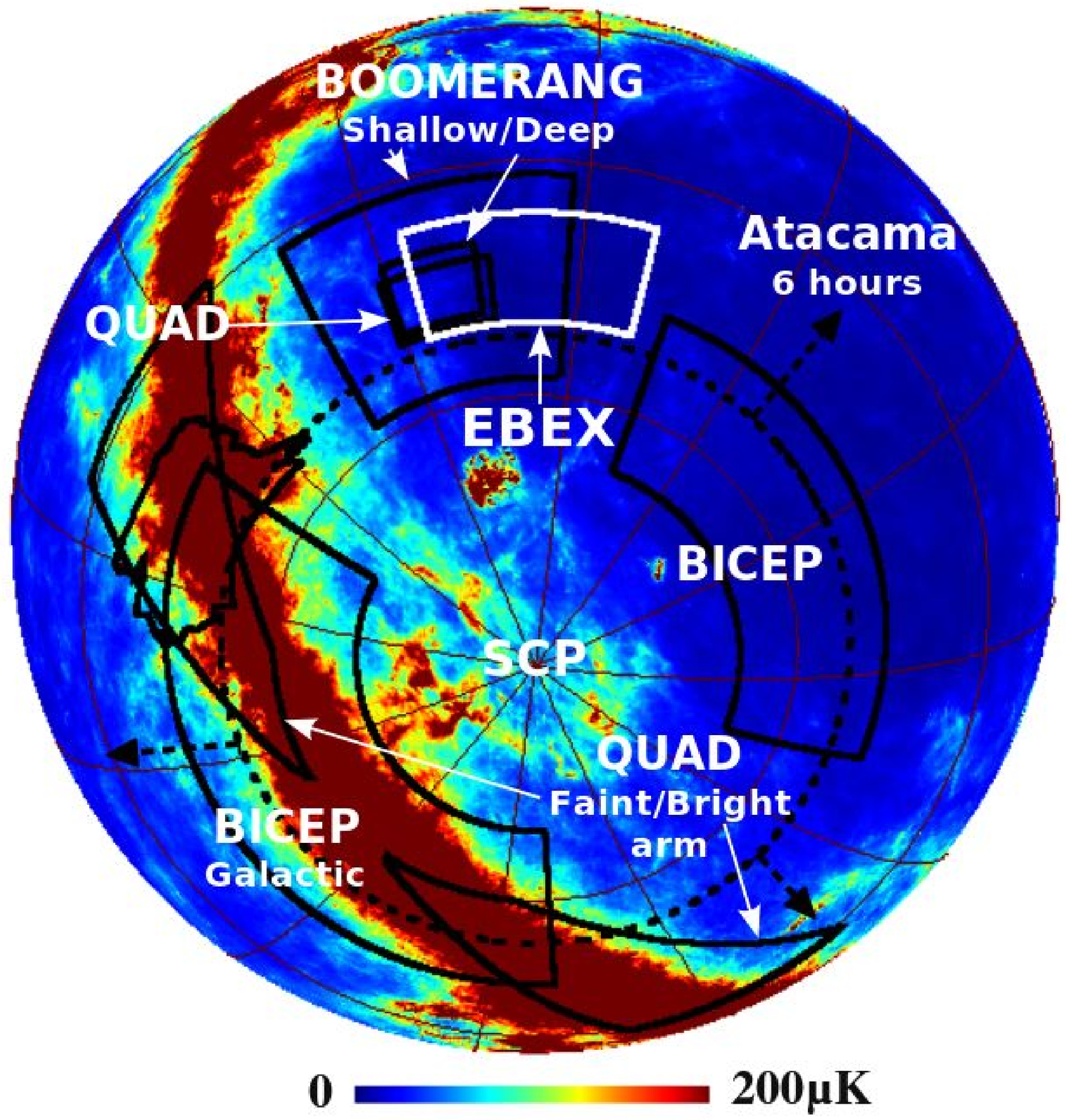}
}
\hspace{0cm}
\subfigure[] % caption for subfigure b
{
    \label{skycov:sub:b}
    \includegraphics[height=1.9in]{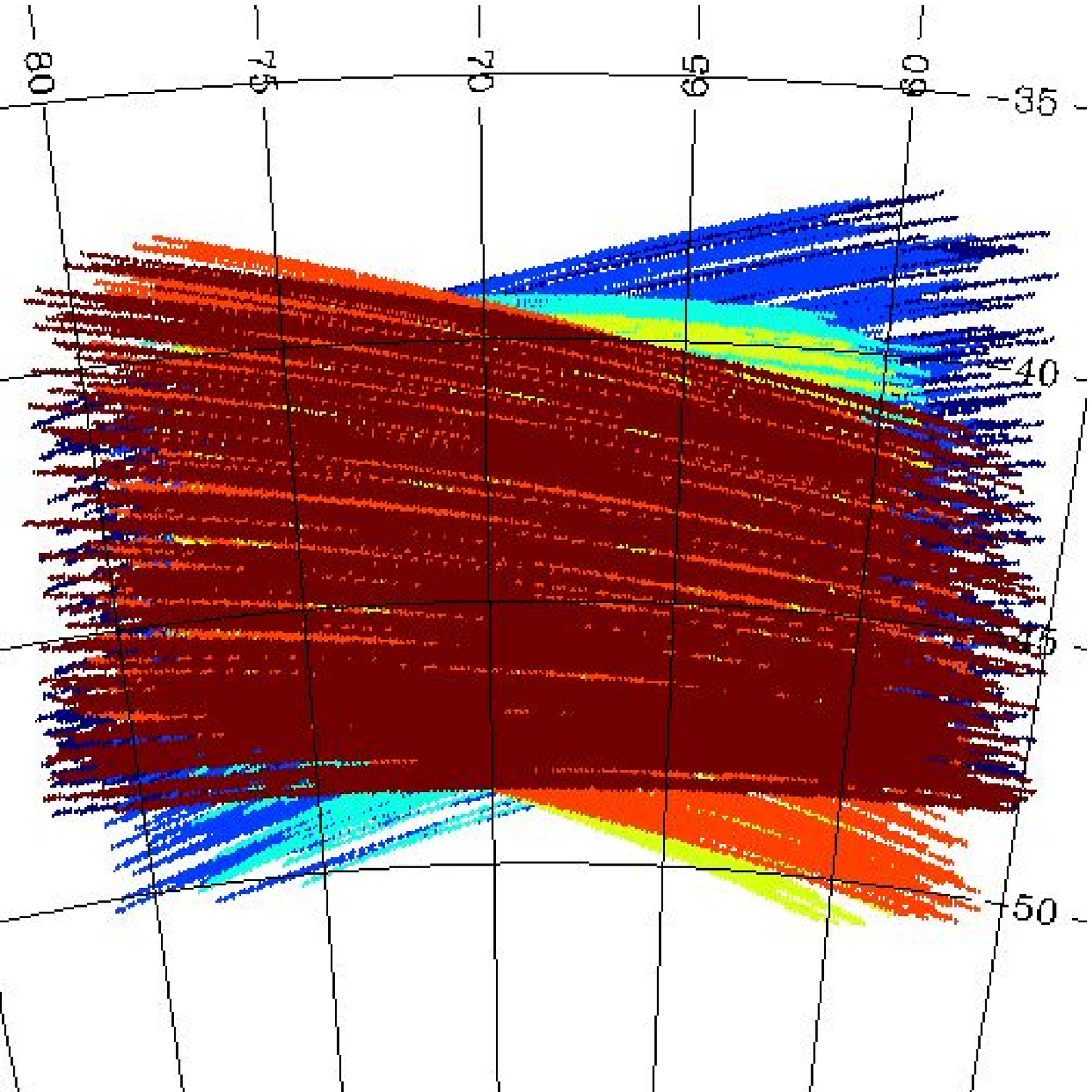}
}
\label{coverage}
\caption[The EBEX scan and sky coverage]{\textit{a:}  The EBEX CMB patch in the southern sky shown outlined in white along with the observing areas of some other CMB experiments.  The linear color scale shows flux at 150 GHz based on Finkbeiner et al.~\cite{Finkbeiner:1999} Model 8, cut off at 200 $\mu K$.  \textit{b:}  24 hours of scans on the sky by a single 150 GHz detector; each color represents a 4 hour scan.} 
\end{figure}

During the long duration science flight over Antartica, EBEX will map a patch near the sourthern celestial pole covering $\sim$ 1\% of the sky, shown in Figure \ref{skycov:sub:a}. Additionally EBEX will spend time scanning polarized and unpolarized calibration sources.  Candidate calibration sources that will be accessible during the Antarctic flight include RCW38, a compact HII region, and Centaurus A, a nearby galaxy with an active nucleus. While mapping the EBEX CMB patch the instrument will perform constant elevation azimuth slews with a 20 degree peak-to-peak amplitude at about 0.4 deg/s, allowing for approximately four measurements of Q and U per 8' beam. After several scan periods the instrument will step up in elevation, with the new central azimuth adjusted to maintain a constant RA. After about four hours the scan will be repeated, where the starting elevation will be adjusted to match the starting declination of the previous scan.  Figure \ref{skycov:sub:b} shows the repeated four hour scans of a single 150 GHz detector over the course of one day.

\section{CONTROL OF SYSTEMATIC EFFECTS}
\label{systematics}
Given the relatively small size of the lensing and gravitational wave $B$-mode signals, control of polarized systematic effects is essential to limit the leakage of $E$ into $B$ (from cross polarization), and of T into $E$ and $B$ (from instrumental polarization).  Systematic effects in EBEX are minimized both in instrument design and in analysis, discussed below in this section.  Calculations based on guidelines from Hu et al.~\cite{Hu:2003b} show that the characterizations required for the absolute cross polarization and the instrumental polarization to insure that these systematic errors contribute negligibly to the final signal are 0.3\% and 0.05\%, respectively.  The sidelobes have already been measured to the required level of 85 dB.  

The use of a HWP (rotating at a frequency, f) provides strong mitigation against a large number of systematic effects.  The CMB polarization signal generated by HWP modulating azimuth scans on the sky will appear in the sidebands of the 4f modulation signal, above the 1/f knee in the detector noise spectrum.  Signals resulting from systematic effects generated by cryostat components downstream of the HWP or any signal stable on scan timescales will be rejected by the demodulation process since these signals will not be changing at 4f.  Additionally, rotating the polarization vector allows for independent measurements of the $I$, $Q$ and $U$ Stokes parameters at each detector during a single pointing at each pixel on the sky, eliminating the need to difference detectors which may have varied gains, absolute calibrations, beams, and noise characteristics.  Any noise sources present at frequencies other than 4f will be rejected by demodulation; signals that may arise from the HWP motor and other rotation hardware will reside at f and 2f.  Also, as discussed above, the 9" reconstruction pointing requirement ensures minimal impact on the data from mixing of $E$ and $B$ due to pointing errors, and the Gregorian Mizuguchi-Dragone optical design minimizes polarized systematic effects.

Systematic errors will be controlled in analysis by differencing maps produced for individual detectors and spatial and temporal jack-knives in the map domain.  Additionally, mixing of $E$ and $B$ due to the presence of ambiguous modes in the partial sky coverage case is limited by using an estimator for $E$ and $B$ developed by Smith and Zaldarriaga~\cite{Smith:2007b}.  Finally, the opposite parity of $E$ and $B$ can be exploited to constrain and remove systematic effects, such as rotations in the focal plane, assuming no parity violating processes in the early universe~\cite{Kamionkowski:2009, Yadav:2009}.  The measured $EB$ cross-correlation is assumed to be of non-primordial origin, resulting from either instrumental effects or some astrophysical process since decoupling.  Simulations show that the overall rotation of the focal plane can be recovered, with the variance of the recovered rotation increasing with input rotation angle; for rotation angles of a few degrees or less, the recovered angle can be constrained to an order of magnitude better than that required by EBEX.  The relative rotation angle of detectors can also be recovered.  Additionally, simulations show that the rotation operation can be inverted to recover a de-rotated $BB$ power spectrum for small rotation angles.  The simulations show that no significant $EB$ correlation is generated by boundary effects on the observation patch edges since the $EB$ correlation oscillates around zero.
%The instrumental polarization is expected to be dominated by the anti-reflection (AR) coating on the first lens in the cryostat, at a level of 0.5 to 4\% depending on the final implementation of the coating, and the HWP.  However, since this signal will be stable on scan timescales modulation with the HWP will minimize its presence in the demodulated signal.  
%The EBEX cross polarization will be measured on the ground before flight to 0.2\% at each detector, and to 0.03\% differentially between detector pairs, better than the absolute characterization requirement of 0.3\% set by the requirement that these errors contribute negligibly to the final signal.  Ground measurements will be confirmed with flight data, where absolute rotation can be constrained to 0.04\% by exploiting the opposite parities of $E$ and B, requiring the the $EB$ cross correlation to vanish.  The instrumental polarization needs to be characterized to 0.05\%; calculations show that characterization and subtraction can be achieved to 0.01\% using observations of the CMB dipole in the EBEX CMB patch.
\section{REMOVAL OF FOREGROUND CONTAMINANTS}
\label{sec:foregrounds}
Although EBEX will observe in a region with an exceptionally low amplitude of polarized foreground emission, efficient subtraction of the polarized dust foreground from the raw data will be necessary to measure both the gravitational wave and lensing $B$-modes, as shown in Figure \ref{sci:sub:b}.  EBEX will employ a parametric component separation technique~\cite{Eriksen:2006,Stompor:2009} to reconstruct the power spectrum of the dust foreground and to clean the dust signal from the CMB.  The technique exploits the distinct frequency scaling of the CMB and dust signals.  

Stivoli et al.~\cite{Stivoli:2010} investigate the performance of the maximum likelihood parametric method on simulated data for
a balloon-borne experiment similar to EBEX in frequency coverage, sensitivity, scan area and patch location.  They
study the consequences of the component separation step on an estimation of the $B$-mode power spectrum and $r$.  In their model the dust and CMB in every pixel are characterized by unknown amplitudes. The spectral index for dust is also unknown. Their formalism was used to estimate the impact of foreground subtraction for EBEX. They find that the major impact on the determination of $r$ comes from the effects of instrument noise on assessing the foreground level and in producing cleaned CMB maps. The effect of the error in the determination of the dust spectral index is typically sub-dominant. For this reason the values quoted in Section~\ref{sec:probeinfl} for the area of the likelihood function neglect the uncertainty in the determination of the spectral index.  Stivoli et al. also show that these results are robust with respect to the presence of small discrepancies between the data and model applied on the component separation step, as allowed for by the current knowledge of the foregrounds in the considered sky area or due to some instrumental errors,
such as relative calibration and/or frequency channel band uncertainties.

%Stivoli et al., 2010~\cite{Stivoli:2010} implements a maximum likelihood method to simulated data for a balloon-borne experiment similar to EBEX in frequency coverage, sensitivity, scan area and patch location.  In the simulations they allow for a single dust spectral index, additional dust power at small scales, and inclusion of a synchrotron component.  They also include simulations with calibration errors of the high frequency channels relative to the 150 GHz band and deviations of the power spectrum from a single spectral index.  They find that, assuming the expected calibration uncertainties for EBEX and values for the spectral mismatch, the prescribed method of foreground subtraction will allow for a measurement of r down to 0.04 at 95\% confidence, including sample variance, noise, E to $B$ leakage, and foreground residuals\footnote[1]{The actual value of r which can be measured by EBEX, 0.1, is determined by the instrument sensitivity.}.  

\section{OVERVIEW OF THE NORTH AMERICAN ENGINEERING FLIGHT}

During the engineering flight the receiver contained one focal plane with one wafer at each observation frequency; 64, 32, and 71 detectors at 150, 250, and 410 GHz, respectively, were were exposed to radiation through the cryostat window and read out.  The lenses and HWP had no AR coating and neutral density filters made from MF-110 {ECCOSORB}\textregistered{} were placed in front of the 250 and 410 GHz wafers to allow for receiver testing on the ground.  Additionally, some detectors were covered with {ECCOSORB}\textregistered{} plugs or made dark with aluminum tape.  EBEX was launched from the CSBF facility at Ft. Sumner, NM, on June 11, 2009, and it drifted at an altitude of about  35 km (115 kft) to the far western border of Arizona in about 14 hours.  The altitude and air temperature were generally stable and daytime conditions were similar to those expected during the long duration flight.  

Milestones from the flight include the first implementation of millimeter-wave TES bolometers and a SQuID-based multiplexed readout operated in a space-like environment, where the detectors and readout system were tuned end-to-end remotely from the ground, and the first use of a broadband HWP rotating on a SMB, which turned continuously during entire flight.  Additionally, the cryogenics functioned as expected maintaining detectors at 0.27 K, the temperatures of all systems were maintained within the specified operating range, continuous command uplink and data downlink telemetry was maintained throughout flight, and CMB patch and dipole scans were performed.  A detailed discussion of the flight performance of the detectors and readout electronics is provided elsewhere in these proceedings~\cite{Aubin:2010}.  The performance of the cryogenics and receiver housekeeping during the flight is also available elsewhere~\cite{Sagiv:2010}.

\section{STATUS}

We are currently analyzing data from the engineering flight and preparing the instrument for the long duration flight.  A second focal plane is being added to the receiver and both focal planes will be fully populated with detector wafers.  Anti-reflection coatings will be added to the lenses and HWP to improve the receiver efficiency.  The solar power system is being built and the second redundant star camera and gyroscope box will be integrated into the attitude control system.  The inner and outer frames of the gondola and the cryostat are being modified to reduce the payload weight and to allow for dynamical balance of the inner frame.

\acknowledgments     %>>>> equivalent to \section*{ACKNOWLEDGMENTS}       
 
EBEX is a NASA supported mission through grant numbers NNX08AG40G and NNX07AP36H. We thank Columbia ScientiÞc Balloon Facility for their enthusiastic support of EBEX. We also acknowledge support from CNRS, Minnesota Super Computing Institute and the Science and Technology Facilities Council. This research used resources of the National Energy Research ScientiÞc Computing Center, which is supported by the office of Science of the U.S. Department of Energy under contract No. DE-AC02-05CH11231. The McGill authors acknowledge funding from the Canadian Space Agency, Natural Sciences and Engineering Research Council, Canadian Institute for Advanced Research, Canadian Foundation for Innovation and Canada Research Chairs program. 

%%%%%%%%%%%%%%%%%%%%%%%%%%%%%%%%%%%%%%%%%%%%%%%%%%%%%%%%%%%%%
%%%%% References %%%%%

\bibliography{spiemanusc_brk_etal}   %>>>> bibliography data in report.bib
\bibliographystyle{spiebib}   %>>>> makes bibtex use spiebib.bst

\end{document}